\documentclass[10pt]{article}
\usepackage{epsfig}
\renewcommand{\author}[2]{\begin{center}
                           {\sc #1}\\
                           {#2}
                          \end{center}}
\renewcommand{\title}[1]{\begin{center}
                      {\Large {\bf #1}}
                      \end{center}}

 \def\etal{{\sl et al.}\,}
 \def\ie{{\sl i.e.,}\,}
 \def\eg{{\sl e.g.,}\,}

 \def\la{\hbox{\raise.5ex\hbox{$<$}
     \kern-1.1em\lower.5ex\hbox{$\sim$}}}
 \def\ga{\hbox{\raise.5ex\hbox{$>$}
     \kern-1.1em\lower.5ex\hbox{$\sim$}}}
 \def\sun{\odot}
 \def\ms{M_{\sun}}

\def\apj{{\it ApJ}}

\def\apjs{{\it ApJS}}

\def\aap{{\it A\&A}}

\def\mnras{{\it MNRAS}}

\def\nat{{\it Nature}}

\begin{document}
\title{Relativistic Jets from Collapsars
\vspace*{4mm}}

\author{M.A. Aloy$^1$, 
        E. M\"uller$^2$,
        J.M$^{\underline{\mbox{a}}}$ Ib\'a\~nez$^1$,
        J.M$^{\underline{\mbox{a}}}$ Mart\'{\i}$^1$ and
        A. MacFadyen$^3$
\vspace{2mm}}

\begin{center}
\small{$^1$Departamento de Astronom\'{\i}a y Astrof\'{\i}sica,
                  UVEG, 46100 Burjassot,
                  Spain. \\ e-mails:~Miguel.A.Aloy@uv.es,
                  Jose.M.Ibanez@uv.es, Jose.M.Marti@uv.es}
\vspace{2mm}

\noindent
\small{$^2$Max-Planck-Institut f\"ur Astrophysik,
           85748 Garching, Germany. \\ e-mail:~ewald@mpa-garching.mpg.de}
\vspace{2mm}

\noindent
\small{$^3$Astronomy Department, University of California, Santa Cruz, 
           CA 95064. \\ e-mail:~andrew@ucolick.org}
\end{center}

\vspace{3mm} 

\begin{abstract}
We have studied the relativistic beamed outflow proposed to occur in
the collapsar model of gamma--ray bursts. A jet forms as a consequence
of an assumed energy deposition of $\sim 10^{50}- 10^{51}$\,erg/s
within a $30^{\circ}$ cone around the rotation axis of the progenitor
star.  The generated jet flow is strongly beamed ($\la$ few degrees)
and reaches the surface of the stellar progenitor (r $\approx
3\,10^{10}\,$cm) intact.  At break-out the maximum Lorentz factor of
the jet flow is about 33. Simulations have been performed with the
GENESIS multi-dimensional relativistic hydrodynamic code.
\end{abstract}

\section{Motivation and numerical setup} \label{s:intro}

Various catastrophic collapse events have been proposed to explain the
energies released in a gamma--ray burst (GRB) including compact binary
system mergers \cite{Go86, MH93}, collapsars \cite{Wo93}
and hypernovae \cite{Pa98}. These models all rely on a common engine,
namely a stellar mass black hole (BH) which accretes several solar
masses of matter from a disk (formed during a merger or by a
non--spherical collapse). A fraction of the gravitational binding
energy released by accretion is converted into a pair
fireball. Provided the baryon load of the fireball is not too large,
the baryons are accelerated together with the e$^+\,$e$^-$ pairs to
ultra--relativistic speeds (Lorentz factors $> 10^2$;
\cite{CR78}). The existence of such relativistic flows is supported by
radio observations of GRB\,980425 \cite{KF98}.

The dynamics of spherically symmetric relativistic fireballs has been
studied by several authors by means of 1D Lagrangian hydrodynamic
simulations (\eg \cite{ML93}). It has been
argued that the rapid temporal decay of several GRB afterglows is more
consistent with the evolution of a relativistic jet after it slows
down and spreads laterally than with a spherical blast wave
\cite{KD99}.  The lack of a significant radio afterglow in
GRB\,990123 provides independent evidence for jet--like geometry
\cite{KF99}. Motivated by these observations and by the collapsar
model of \cite{MW99}, we have simulated the propagation of jets from
collapsars using relativistic hydrodynamics.

In \cite{MW99} the continued evolution of rotating helium stars, whose
iron core collapse does not produce a successful outgoing shock but
instead forms a BH surrounded by a compact accretion disk, has been
explored. Assuming that the efficiency of energy deposition by $\nu
\bar\nu$--annihilation or, \eg magneto-hydrodynamic processes is
higher in the polar regions, \cite{MW99} obtained relativistic jets
along the rotation axis, which remained highly focused, and capable of
penetrating the star.  However, as these simulations were performed
with a Newtonian hydrodynamic code, appreciably superluminal speeds in
the jet flow were obtained.

We have performed axisymmetric relativistic simulations of jets from
collapsars starting from Model\,14A of \cite{MW99}.  The simulations
have been performed with GENESIS a multidimensional relativistic
hydrodynamic code (based on Godunov-type schemes) developed by
\cite{AI99} using 2D spherical coordinates ($r, \theta$).  GENESIS
employs a 3th order explicit Runge--Kutta method \cite{SO89} to
advance in the time the relativistic Euler equations written in
conservation form. High spatial order is provided by a PPM
reconstruction \cite{CW84} that sets up the values of the physical
variables in order to solve linearized Riemann problems at every cell
interface (using the Marquina's flux formula \cite{DF98}).

The innermost $2.03\,\ms$ representing the iron core were removed from
the helium star model by introducing an inner boundary at a radius of
$200\,$km.  When the central BH has acquired a mass of $3.762\,\ms$,
we mapped the model to our computational grid.  In the $r$--direction
the computational grid consists of 200 zones spaced logarithmically
between the inner boundary and the surface of the helium star at
$R_{*} = 2.98\times 10^{10}\,$cm. Assuming equatorial--plane symmetry
we use four different zonings in the angular direction: 44, 90 and 180
uniform zones (\ie $2^{\circ}, 1^{\circ}$ and $0.5^{\circ}$ angular
resolution), and 100 nonuniform zones covering the polar region
$0^{\circ} \le \theta \le 30^{\circ}$ with 60 equidistant zones
($0.5^{\circ}$ resolution) and the remaining 40 zones being
logarithmically distributed between $30^{\circ} \le \theta \le
90^{\circ}$.

The gravitational field of the BH is described by the static
Schwarzschild metric, neglecting the effects due to self--gravity of
the star. We used the EOS of \cite{WJ94} which includes the
contribution of non--relativistic nucleons treated as a mixture of
Boltzmann gases, and radiation, as well as an approximate correction
due to pairs $e^+e^-$. Full ionization and non-degeneracy of the
electrons is assumed. We advect (\ie we do not solve additional
Riemann problems for each component) nine non-reacting nuclear species
which are present in the initial model.

In a consistent collapsar model the jet will be launched by any
physical process which gives rise to a local deposition of energy
and/or momentum. We mimic this process by depositing energy at a
constant rate, $\dot E$, within a $30^{\circ}$ cone around the
rotation axis of the progenitor star. In radial direction the
deposition region extends from the inner boundary to a radius of
$6\times10^7\,$cm.  We consider two cases that bracket the expected
$\dot E$ of the collapsar models: $10^{50}\,$erg/s, and $10^{51}\,$erg/s.

\section{Results} 
\label{s:results}

{\bf Low energy deposition rate (Model A).}  Using a constant
$\dot E = 10^{50}\,$erg/s a relativistic jet forms within a fraction
of a second and starts to propagate along the rotation axis
(Fig.\,1). The jet exhibits all the typical morphological elements
\cite{BR74}: a terminal bow shock, a narrow cocoon, a contact
discontinuity separating ste-%
\begin{figure}[hb]
\centerline{\epsfig{file=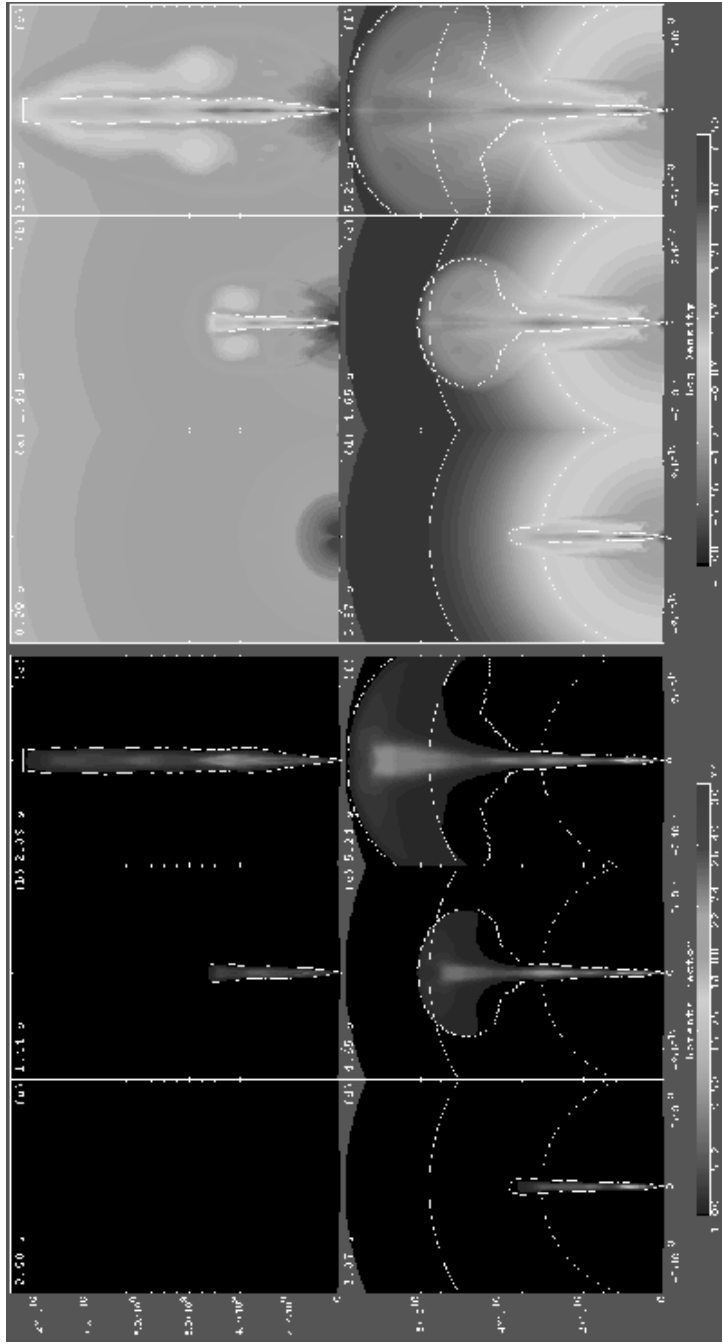,width=9.5cm}}
\caption{\small Coloured contour maps of the logarithm of the
rest--mass density (six top panels) and the Lorentz factor for model A
at different evolution times. Note the change in the scale between
left and right panels.
\label{f:lorerho}}
\end{figure}
\clearpage
\noindent
llar and jet matter, and a hot spot.  The propagation of the
jet is unsteady, because of density inhomogeneities in the star. The
Lorentz factor of the jet, $W$, increases non--monotonically with
time, while the density drops to $\sim 10^{-6}$\,gr/cm$^3$. The
density profile shows large variations (up to a factor of 100) due to
internal shocks. The mean density in the jet is $\sim 10^{-2} - 1$
\,g/cm$^3$.

Some of the internal shocks are biconical and recollimate the
beam. These shocks develop during the jet's propagation and may
provide the ``internal shocks'' proposed to explain the observed
gamma--ray emission \cite{Ka94}. A particularly strong recollimation
shock forms during the early stages of the evolution, followed by a
strong rarefaction that causes the largest acceleration of the beam
material giving rise to a maximum in $W$.  When the jet encounters a
region along the axis where the density gradient is positive the jet's
head is decelerated, while the a central channel in the beam is
cleaned by outflow into the cocoon through the head. This leads to an
acceleration of the beam. The combination of both effects
(deceleration of the head and beam acceleration) increases the
strength of the internal shocks.

The relativistic treatment of the hydrodynamics leads to an overall
qualitatively similar evolution than in \cite{MW99} (formation of a
jet), being, however, a quantitatively very different. We find that
the results strongly depend on the angular resolution, and the
minimum acceptable one is $0.5^{\circ}$ (at least near the axis). At
this resolution we find $W_{\rm max} \sim 15-20$ (at
shock break--out) at a radius $\sim 8\times10^9$\,cm. Within the
uncertainties of the jet mass determinations due to finite zoning and
the lack of a precise numerical criterium to identify jet matter, the
baryon load, $\eta$, seems to decrease with increasing resolution. In
the highest resolution run we find $\eta \simeq 1.3 \pm 1.2 $ at shock
break-out (see also Sect.\,4).

{\bf High energy deposition rate (Model B).}  Enhancing $\dot E$
by a tenfold ($\dot E = 10^{51}$\,erg/s), the jet flow reaches larger
values of $W_{\rm max}$. We observe transients during which $W_{\rm
max}$ becomes as large as 40 ($W_{\rm max} =33.3$ at shock
breakout). The jet propagates faster than in model A. The time
required to reach the surface of the star is 2.27\,s instead of
3.35\,s. The opening angle of the jet at shock breakout is $\sim
10^{\circ}$, \ie the jets is less collimated than model A. The
strong recollimation shock present in the model A is not so
evident here. Instead, several biconical shocks are observed, and $W$
near the head of the jet is larger ($\sim 22$ in the final model)
because, due to the larger $\dot E$, the central funnel is evacuated
faster, and because the mean density in the jet is 5 times smaller
than in model A ($\eta$ being twice as large).

\begin{figure}[hbt]
\centerline{\psfig{file=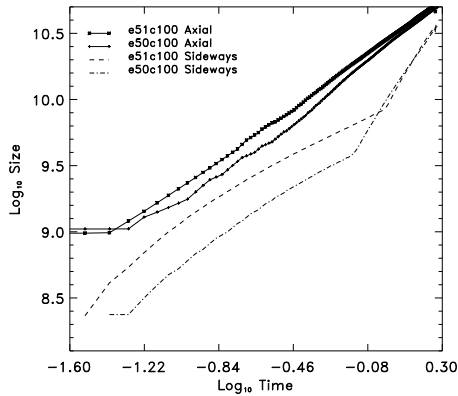,width=6cm}}
\caption{\small Evolution of the axial and lateral sizes of the jet
cavity during the post--breakout epoch. Time is measured 
with respect to the breakout time for each model.
\label{f:size}}
\end{figure}

{\bf Evolution after shock breakout.}
After reaching the stellar surface the relativistic jet propagates
through a medium of decreasing density continuously releasing energy
into a medium whose pressure is negligible compared to that in the jet
cavity, and whose density is (initially) of the same order as that of
the jet. These are jump conditions that generate a strong blast wave.
The external density gradient determines whether the shock will
accelerate or decelerate with time (\cite{Sh79}). In order to satisfy
the conditions for accelerating shocks (\cite{Sh79}), we have
generated a Gaussian atmosphere matching an external uniform medium.
We use models A and B to simulate the evolution after
shock breakout. The computational domain is extended for this purpose
to a radius of $R_t = 7.6\times10^{10}$\,cm. The jet reaches $R_t$
(from the stellar surface) after 1.8\,s in both models, \ie the mean
propagation velocity is $\sim 0.85c$ (almost three times larger than
that inside the star).

The evolution after shock breakout can be distinguished into three
epochs (see Figs.\,\ref{f:lorerho} and \ref{f:size}), which are
related with (i) the external thermodynamical gradients and (ii) the
importance of the axial momentum flux relative to the pressure into
the jet cavity. Both effects determine the shape of the expanding
bubble --prolate-- (see Figs.\,\ref{f:lorerho} and \ref{f:size})
during the post--breakout evolution. However, when the jet reaches the
uniform part of the circumstellar environment, the shape changes
appreciably, because the sideways expansion is faster.  We have not
followed the evolution long enough to see what happens when most of
the bubble has reached the uniform part of the
environment. Nevertheless, we can infer from Fig.\,2 than the widening
rate reduces with time in a way similar to what has happened to the
axial expansion. At latter times most of the bubble is inside the
uniform medium, and the bubble will eventually be pressure
driven. Hence a isotropic expansion is expected.

After shock breakout there are transients in which $W_{\rm max}$
becomes almost 50 in some parts of the beam, $W_{\rm max}$ is again
obtained behind the strongest recollimation shock. The Lorentz factor
near the boundary of the cavity blown by the jet grows from $\sim1$
(at shock breakout) to $\sim 3$ in both models decreasing with
latitude. At the end of the simulation $W_{\rm max}$ is 29.35 (44.17)
for model A (B), which is still smaller than the ones
required for the fireball model (\cite{CR78}).  However, our
simulations have not been pushed far enough in time yet and, therfore,
they can (at the present stage) neither account for the observational
properties of GRBs nor of their afterglows. Instead, our set of
numerical models can be regarded as simulations of a proto-GRB,
because the scales treated in the simulations are still by more than
100 times smaller than the typical distances at which the fireball
eventually becomes optically thin ($\sim 10^{13}$\,cm).


\begin{thebibliography}{}
%
\bibitem{AI99} Aloy, M.A., Ib\'a\~nez, J.M$^{\underline{\mbox{a}}}$,
               Mart\'{\i}, J.M$^{\underline{\mbox{a}}}$ and M\"uller,
               E. (1999), GENESIS: A high-resolution code for
               3D relativistic hydrodynamics. \apjs,
               {\bf 122}, 151.
%
\bibitem{BR74} Blandford, R.D., \& Rees, M.J (1974), A ``twin-exhaust''
               model for double radio sources. \mnras, {\bf 169}, 395.
%
\bibitem{CR78} Cavallo, G. \& Rees, M.J. (1978), A qualitative study
               of cosmic fireballs and gamma-ray bursts. \mnras, {\bf 183},
               359.
%
\bibitem{CW84} Colella, P., \& Woodward, P.R. 1984, The Piecewise
               Parabolic Method (PPM) for Gas-Dynamical Simulations. 
               {\it JCP}, {\bf 54}, 174.
%
\bibitem{DF98} Donat, R., Font, J.A., Ib\'a\~nez,
               J.M$^{\underline{\mbox{a}}}$., \& Marquina, A. (1998), A 
               Flux-Split Algorithm Applied to Relativistic Flows. {\it JCP}, 
               {\bf 146}, 58.
%
\bibitem{Go86} Goodman, J. (1986), Are gamma-ray bursts optically
               thick?. \apj, {\bf 308}, L47.
%
\bibitem{Ka94} Katz, J.I. (1994), Two populations and models of gamma-ray 
               bursts \apj, {\bf 422}, 248.
%
\bibitem{KF98} Kulkarni, S.R., \etal (1998), Radio emission from the
	       unusual SN1998bw and its association with the
	       gamma-ray burst of 25 April 1998. \nat, {\bf 395}, 663.
%
\bibitem{KD99} Kulkarni, S.R., \etal (1999a), The afterglow, redshift
               and extreme energetics of the gamma-ray burst of 23
               January 1999. \nat, {\bf 398}, 389.
%
\bibitem{KF99} Kulkarni, S.R., \etal (1999b), Discovery of a Radio
               Flare from GRB 990123. \apj, {\bf 522}, 97.
%
\bibitem{MW99} MacFadyen, A. and Woosley, S.E. (1999), Collapsars -
               Gamma-Ray Bursts and Explosions in "Failed Supernovae"
               \apj, in press; and astro-ph/9810274.
%
\bibitem{ML93} M\`ez\'aros, P., Laguna, P., \& Rees, M.J. (1993),
 	       Gasdynamics of relativistically expanding gamma-ray
 	       burst sources. \apj, {\bf 415}, 181.
%
\bibitem{MH93} Mochkovitch, R., Hernanz, M., Isern, J., \& Martin, X.
               (1993), GRBs as collimated jets from
               neutron star/black hole mergers. \nat, {\bf 361}, 236.
%
\bibitem{Pa98} Pacy\'nski, B. (1998), Are GRBs in
	       Star-Forming Regions?. \apj, {\bf 494}, L45.
%
\bibitem{Sh79} Shapiro, P.R. (1979), Relativistic blast waves in 2D. 
               \apj, {\bf 233}, 831.
%
\bibitem{SO89} Shu, C.W., \& Osher, S.J. (1989), Efficient Implementation of 
               Essentially Non-Oscillatory shock-capturing schemes. 2.
               {\it JCP}, {\bf 83}, 32.
%
\bibitem{WJ94} Witti, Janka, H.-T., \& Takahashi, (1994),
               Nucleosynthesis in neutrino-driven winds from
               protoneutron stars I. The {alpha}-process. \aap, {\bf
               286}, 841.
%
\bibitem{Wo93} Woosley, S.E. (1993), GRBs from stellar mass accretion
	       disks around black holes. \apj, {\bf 405}, 273.
%
\end{thebibliography}
\end{document}